\documentclass[preprint,12pt]{elsarticle}

\usepackage{amssymb}
\usepackage{multirow}
\usepackage{booktabs}
\usepackage{amssymb}
\usepackage{mathrsfs}
\usepackage{longtable}
\usepackage{lscape}
\usepackage{tabularx}
\usepackage{epsfig}
\usepackage{colortbl}
\usepackage{subfigure}
\journal{Elsevier}

\begin{document}

\begin{frontmatter}



\title{Measure the similarity of nodes in the complex networks}


\author[swu]{Qi Zhang}
\author[swu]{Meizhu Li}
\author[swu,NWPU,vu]{Yong Deng\corref{cor}}
\ead{ydeng@swu.edu.cn, prof.deng@hotmail.com}
\author[vu]{Sankaran Mahadevan}

\cortext[cor]{Corresponding author: Yong Deng, School of Computer and Information Science, Southwest University, Chongqing, 400715, China.}

\address[swu]{School of Computer and Information Science, Southwest University, Chongqing, 400715, China}
\address[NWPU]{School of Automation, Northwestern Polytechnical University, Xian, Shaanxi 710072, China}
\address[vu]{School of Engineering, Vanderbilt University, Nashville, TN, 37235, USA}

\begin{abstract}
    Measure the similarity of the nodes in the complex networks have interested many researchers to explore it. In this paper, a new method which is based on the degree centrality and the Relative-entropy is proposed to measure the similarity of the nodes in the complex networks. The results in this paper show that, the nodes which have a common structure property always have a high similarity to others nodes. The nodes which have a high influential to others always have a small value of similarity to other nodes and the marginal nodes also have a low similar to other nodes. The results in this paper show that the proposed method is useful and reasonable to measure the similarity of the nodes in the complex networks.

\end{abstract}
\end{frontmatter}

\section{Introduction}
\label{Introduction}
The complex networks is a new method to describe those complex system from the mathematic. Many of the real system in the real world can be modeled as the complex system, such as the biological, social and technological systems \cite{albert2000error,newman2003structure,de2014facebook,csermely2008creative,csermely2009weak}. Many property of the complex networks have illuminated by these researchers in this filed, such as the network topology and dynamics \cite{watts1998collective,newman2006structure,ferrara2013traveling,ferrara2012large}, the property of the network structure \cite{newman2003structure,barthelemy2004betweenness}, the self-similarity and fractal property of the complex networks\cite{song2005self,wei2014new,zhang2015tsallis}, the evolutionary games on complex networks \cite{wang2013impact,PhysRevE.89.052813}, the controllability and the synchronization of the complex networks \cite{liu2011controllability,arenas2008synchronization} and so on \cite{barabasi2009scale,barabasi1999emergence,barabasi2009scale,meo2013analyzing,ferrara2013traveling,teixeira2010complex,csermely2004strong,wang2012evolution}.

The similarity of the nodes in the complex networks is a new research direction. It is interested that "How similar are these two vertices ?" or " Which node is most similar to others nodes?". There are many methods have proposed to solve this problem \cite{leicht2006vertex,zhou2009predicting,pan2010detecting,lu2001node,lu2007node}. In this paper, a new methods which is based on the relative-entropy (Kullback¨CLeibler divergence) \cite{kullback1951information} is proposed to describe the similarity of those nodes in the complex networks. The definition of the probabilities of each node is based on the degree distribution.

The rest of this paper is organised as follows. Section \ref{Rreparatorywork} introduces some preliminaries of this work. In section \ref{newmethod}, a new method to measure the similarity of the nodes in the complex networks is proposed. The application of the proposed method is illustrated in section \ref{application}. Conclusion is given in Section \ref{conclusion}.
\section{Preliminaries}
\label{Rreparatorywork}

\subsection{Local network in the complex network}
\label{local_network}

Based on the existing research about the complex networks, it is clear that a lot of the property of complex networks are based on the structure property of it \cite{newman2003structure}. In the complex networks, each node's influence on the whole network is decided by the neighbour nodes of it. Based on the existing researches about the local structure of the complex networks \cite{zhou2009predicting,zhou2009predicting,ulanowicz1999nutrient}, a local network of each node in the complex networks is proposed \cite{zhang2014local}. The details of the local networks is shown as follows:

\begin{figure}
    \centering
    \subfigure[Network A]{
    \label{Network_A} 
    \centering
    \includegraphics[scale=0.65]{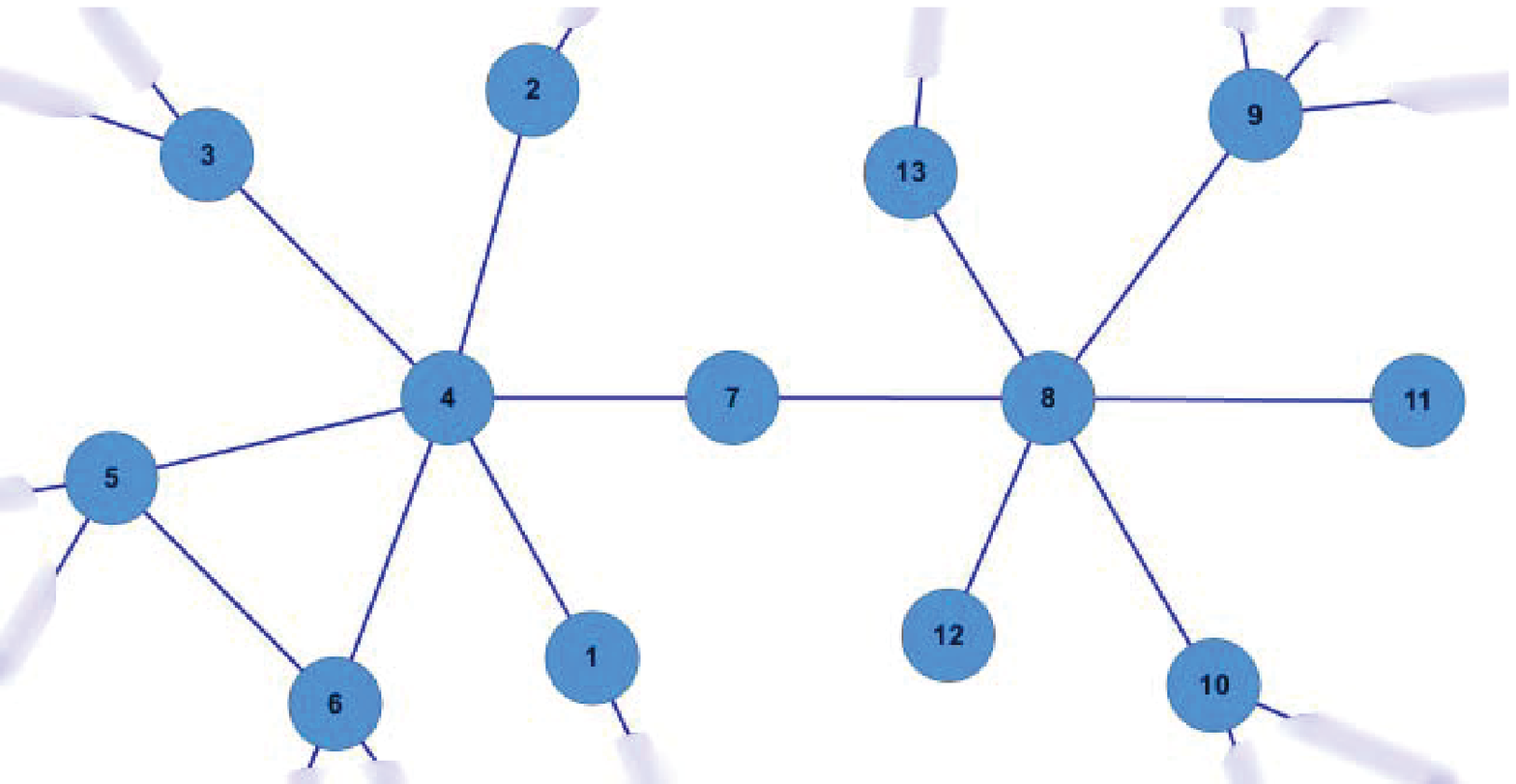}}
    \subfigure[The local network of node 4]{
    \label{4} 
    \centering
    \includegraphics[scale=0.48]{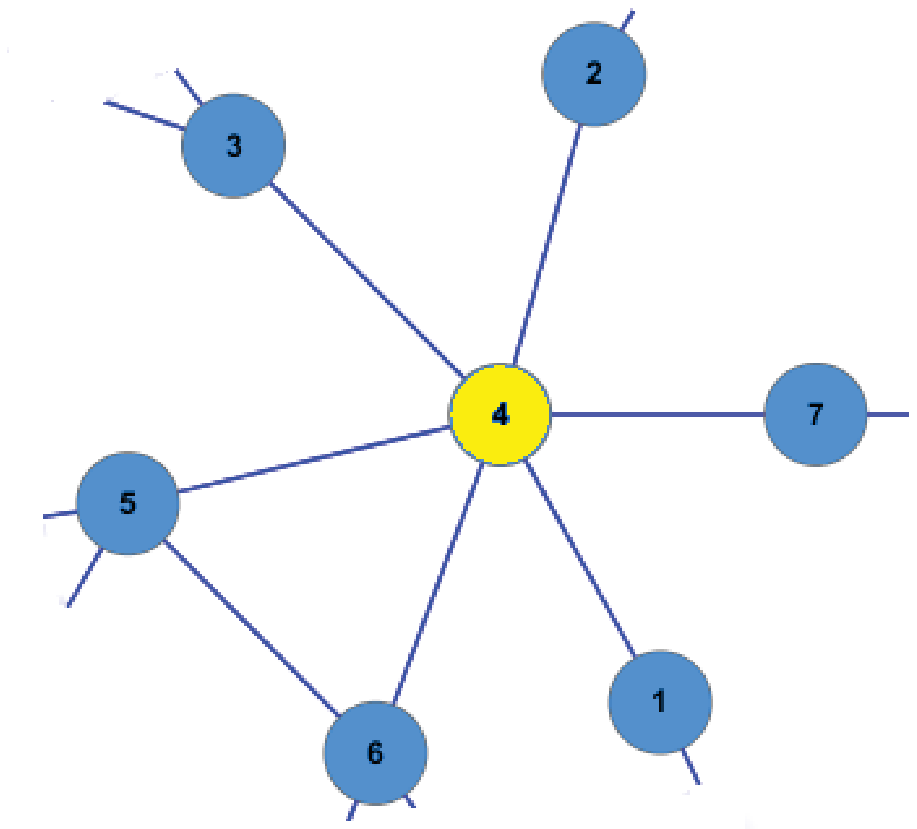}}
    \hspace{0.5cm}
    \subfigure[The local network of node 8]{
    \label{8} 
    \centering
    \includegraphics[scale=0.48]{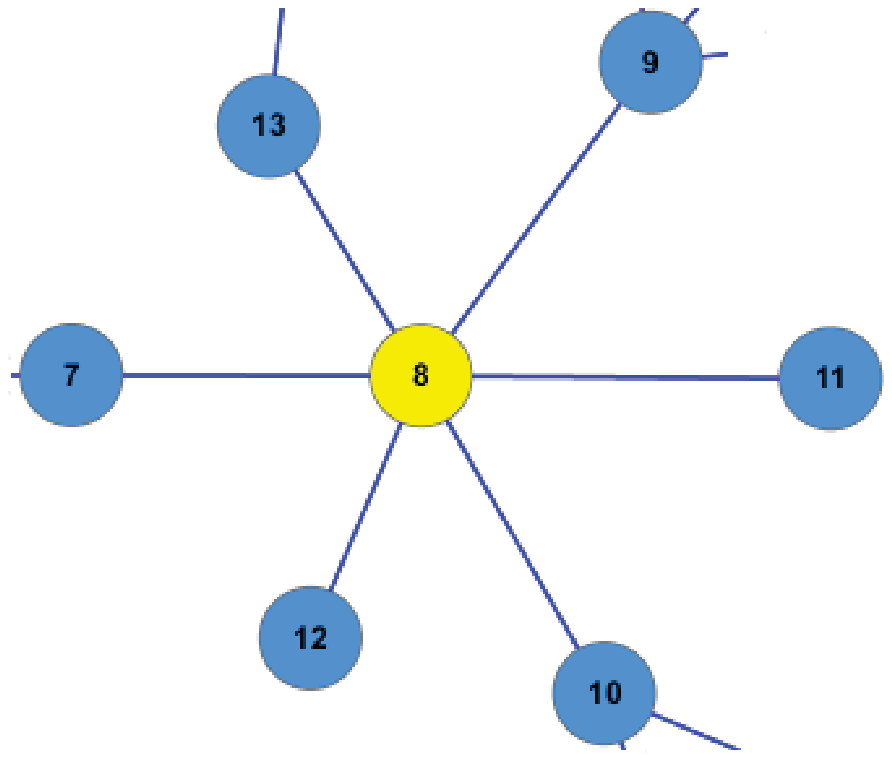}}
  \caption{The Network A in the subfigure (a) is a simple network. The subfigure (b) shows the detail of the local network of node 4. The subfigure (c) shows the details of the local network of node 8.}\label{Local_network}
\end{figure}

It is clear that each local network of the target node contains the target node and the neighbor nodes of the target nodes.

\subsection{Relative entropy (Kullback¨CLeibler divergence)}
\label{Relative_entropy}

The Relative entropy (Kullback¨CLeibler divergence) is a basic conception in the probability theory and the information theory. It is proposed by Kullback and Leibler er.al \cite{kullback1951information}. The Relative entropy is a non-symmetric measure of the difference between two probability. For two probabilities $P$ and $Q$
The definition of the Relative entropy is shown in the Eq.(\ref{D_KL}).

\begin{equation}\label{D_KL}
{D_{KL}}(P||Q) = \sum\limits_{i = 1}^n {P(i)ln\frac{{P(i)}}{{Q(i)}}}
\end{equation}

Where the $q$ and $Q$ have the same number of the components in it. The components in those two probabilities is equal to $n$.
\section{Measure the similarity of each node}
\label{newmethod}
The proposed method is based on the definition of the local network and the Relative entropy. The definition of the proposed new methods can be divided into two parts.

\begin{itemize}
  \item [Part 1] \textbf{The definition of the probabilities of each node}.
        First, calculate the degree of each node. Find the maximum of the degree in the network. Second, set the scales of the probabilities of each node base on the value of the maximum degree. Third, use the degree of the neighbour nodes as the components of probabilities. At last, sort the probabilities from the high to the low.

  \item [Part 2] \textbf{The Relative entropy of each node to others nodes}. Calculate the Relative entropy between each node's probabilities.

\end{itemize}

Based on the local network of each node and the degree centrality, the definition of the probabilities of each node is shown as follows.
For example, we use the $LN(i)$ represents the local network of node $i$.  In the local network $LN(i)$, the total value of degree is represented by the ($T_{degree}(i)$).  The $i$ in the $T_{degree}(i)$ represents the $i$th node. The node number in the local network $LN(i)$ is equal to $k$.
The maximum value of the degree in the whole networks is equal to $max_n$. Then, the number of the components of each node's probabilities is equal to $max_n+1$.
The probabilities of node $i$ is defined in the Eq.(\ref{P_i}).

\begin{equation}\label{P_i}
P(i) = [d(1),d(2),...,d(k),0,0,...,d({max_n+1})]
\end{equation}

where the d(j) in the Eq.(\ref{P_i}) is defined based on the degree of the node in the local network.

In the $P(i)$, the value of $d(i)$ is defined based on the degree of the node in the local network $i$ ($LN(i)$). If the value of node number in the local network is small than the $max_n$, then the value of $d(i)$ will be set as 0. At last, sort the probabilities $P(i)$ from the high to the low.

\begin{figure}[htbp]
  \centering
  \includegraphics[scale=0.6]{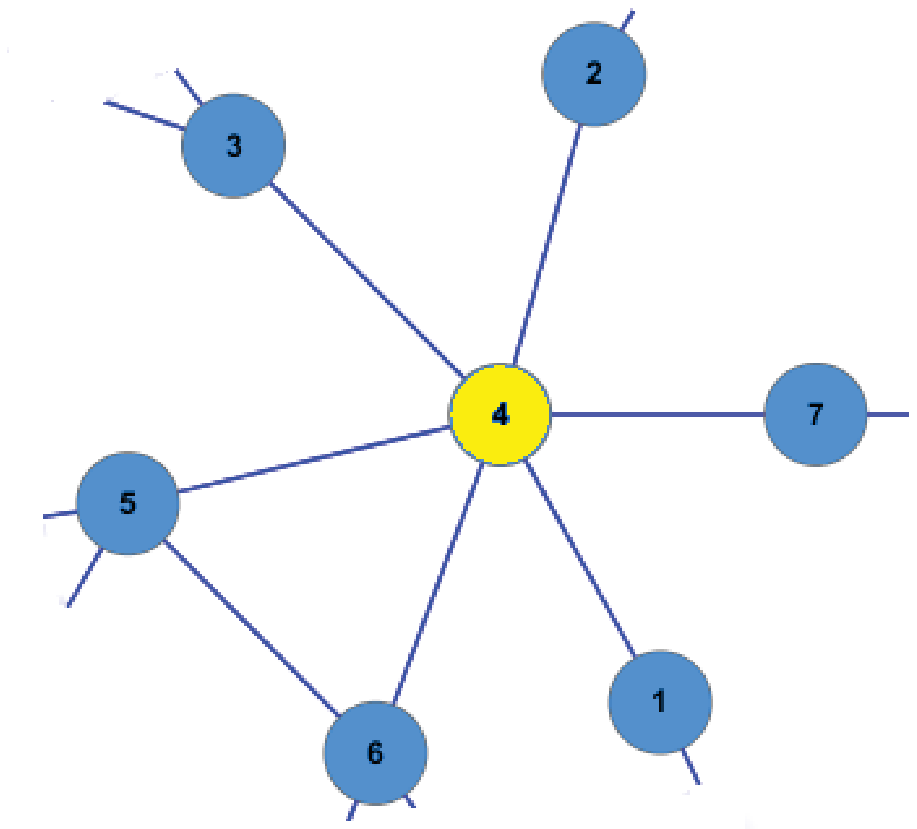}\\
  \caption{This figure is a part of the Network A shows in the Fig.\ref{Local_network}. The maximum of the degree in the Network A is equal to 6. The $max_n$ is the $LN(4)$ is 6. The number of the components in the probabilities is 7.
  The total degree in the $LN(i)$ is equal to 23. The degree of node 1 is 2, the degree of node 2 is 2, the degree of node 3 is 3, the degree of node 5 is 4, the degree of node 6 is 4, the degree of node 7 is 2 and the degree of node 4 is 6. Then $P(4)=[d(1),d(2),d(3),d(4),d(5),d(6),d(7)] = [2/23,2/23,3/23,6/23,4/23,4/23,2/23]$. Then sort the components in the $P(4)$. The $P(4)=[6/23,4/23,4/23,3/23,2/23,2/23,2/23]$.}\label{P_4}
\end{figure}

An example of the definition of $P(i)$ are shown in the Fig. \ref{P_4}.

Then the measure of the similarity of node $i$ and node $j$ is defined as follows:
\begin{equation}\label{m_Simi}
{S_{i,j}} = 1 - ({{{D}}_{KL}}{{(P(i)||P(j))  + }}{{{D}}_{KL}}{{(P(j)||P(i))) }}
\end{equation}
The sum of each node similarity to others in the network is used to identify which node is most similar to others nodes. The big the value of the sum of similarity. The more similar to others nodes.

\begin{figure}[htbp]
  \centering
  \includegraphics[scale=0.6]{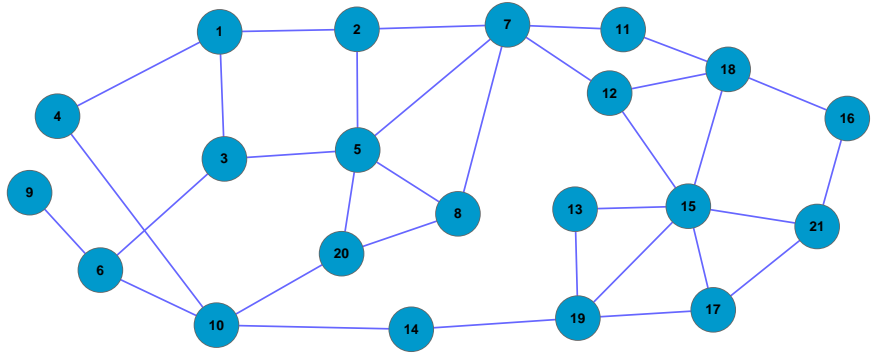}\\
  \caption{The example network (Network A-21)}\label{A_21}
\end{figure}

In order to illuminate the useful of the new method an example network (Network A-21) is used to measure the similarity of nodes in it.
The details of the example network (Network A-21) are shown in the Fig. \ref{A_21}.

The probabilities of each node in the example network (Network A-21) are shown in the Table \ref{tab:probabilities}.
\begin{table}[htbp]
\tiny
\addtolength{\tabcolsep}{-2pt}
  \centering
  \caption{The probabilities ($P(i)$) of each node in the Network A-21}
    \begin{tabular}{cccccccccccccccccccccc}
    \hline
  P(1) = [0.27& 0.27& 0.27& 0.18& 0.00& 0.00& 0.00]  \\
  P(2) = [0.31& 0.31& 0.19& 0.19& 0.00& 0.00& 0.00]  \\
  P(3) = [0.36& 0.21& 0.21& 0.21& 0.00& 0.00& 0.00]  \\
  P(4) = [0.44& 0.33& 0.22& 0.00& 0.00& 0.00& 0.00]  \\
  P(5) = [0.23& 0.23& 0.14& 0.14& 0.14& 0.14& 0.00]  \\
  P(6) = [0.36& 0.27& 0.27& 0.09& 0.00& 0.00& 0.00]  \\
  P(7) = [0.24& 0.24& 0.14& 0.14& 0.14& 0.10& 0.00]  \\
  P(8) = [0.31& 0.31& 0.19& 0.19& 0.00& 0.00& 0.00]  \\
  P(9) = [0.75& 0.25& 0.00& 0.00& 0.00& 0.00& 0.00]  \\
  P(10) = [0.29& 0.21& 0.21& 0.14& 0.14& 0.00& 0.00]  \\
  P(11) = [0.45& 0.36& 0.18& 0.00& 0.00& 0.00& 0.00]  \\
  P(12) = [0.33& 0.28& 0.22& 0.17& 0.00& 0.00& 0.00]  \\
  P(13) = [0.50& 0.33& 0.17& 0.00& 0.00& 0.00& 0.00]  \\
  P(14) = [0.40& 0.40& 0.20& 0.00& 0.00& 0.00& 0.00]  \\
  P(15) = [0.24& 0.16& 0.16& 0.12& 0.12& 0.12& 0.08]  \\
  P(16) = [0.44& 0.33& 0.22& 0.00& 0.00& 0.00& 0.00]  \\
  P(17) = [0.38& 0.25& 0.19& 0.19& 0.00& 0.00& 0.00]  \\
  P(18) = [0.35& 0.24& 0.18& 0.12& 0.12& 0.00& 0.00]  \\
  P(19) = [0.35& 0.24& 0.18& 0.12& 0.12& 0.00& 0.00]  \\
  P(20) = [0.33& 0.27& 0.20& 0.20& 0.00& 0.00& 0.00]  \\
  P(21) = [0.43& 0.21& 0.21& 0.14& 0.00& 0.00& 0.00]  \\
     \hline
    \end{tabular}%
  \label{tab:probabilities}%
\end{table}%

Then the similarity matrix $S_{i,j}$ of the nodes in the example network (Network A-21) is shown in the Eq.(\ref{211}):

\begin{equation}{
\label{211}
{S_{ij}} = \left[
\tiny
\setlength{\arraycolsep}{0.5pt}
  \begin{array}{c|c|c|c|c|c|c|c|c|c|c|c|c|c|c|c|c|c|c|c|c}
    1.00  & 0.96  & 0.94  & 0.89  & 0.88  & 0.91  & 0.90  & 0.96  & 0.52  & 0.96  & 0.84  & 0.98  & 0.80  & 0.88  & 0.85  & 0.89  & 0.93  & 0.90  & 0.90  & 0.96  & 0.89  \\
    0.96  & 1.00  & 0.95  & 0.95  & 0.91  & 0.88  & 0.94  & 1.00  & 0.60  & 0.94  & 0.94  & 0.99  & 0.91  & 0.96  & 0.84  & 0.95  & 0.97  & 0.94  & 0.94  & 0.99  & 0.91  \\
    0.94  & 0.95  & 1.00  & 0.93  & 0.87  & 0.87  & 0.89  & 0.95  & 0.70  & 0.96  & 0.89  & 0.97  & 0.89  & 0.88  & 0.87  & 0.93  & 0.99  & 0.93  & 0.93  & 0.98  & 0.96  \\
    0.89  & 0.95  & 0.93  & 1.00  & 0.77  & 0.96  & 0.80  & 0.95  & 0.82  & 0.88  & 0.99  & 0.96  & 0.98  & 0.98  & 0.73  & 1.00  & 0.96  & 0.93  & 0.93  & 0.95  & 0.95  \\
    0.88  & 0.91  & 0.87  & 0.77  & 1.00  & 0.81  & 0.98  & 0.91  & 0.37  & 0.95  & 0.77  & 0.90  & 0.74  & 0.78  & 0.97  & 0.77  & 0.89  & 0.93  & 0.93  & 0.90  & 0.84  \\
    0.91  & 0.88  & 0.87  & 0.96  & 0.81  & 1.00  & 0.83  & 0.88  & 0.72  & 0.93  & 0.92  & 0.94  & 0.89  & 0.93  & 0.82  & 0.96  & 0.90  & 0.95  & 0.95  & 0.89  & 0.94  \\
    0.90  & 0.94  & 0.89  & 0.80  & 0.98  & 0.83  & 1.00  & 0.94  & 0.41  & 0.96  & 0.80  & 0.92  & 0.77  & 0.81  & 0.95  & 0.80  & 0.91  & 0.94  & 0.94  & 0.93  & 0.86  \\
    0.96  & 1.00  & 0.95  & 0.95  & 0.91  & 0.88  & 0.94  & 1.00  & 0.60  & 0.94  & 0.94  & 0.99  & 0.91  & 0.96  & 0.84  & 0.95  & 0.97  & 0.94  & 0.94  & 0.99  & 0.91  \\
    0.52  & 0.60  & 0.70  & 0.82  & 0.37  & 0.72  & 0.41  & 0.60  & 1.00  & 0.55  & 0.81  & 0.66  & 0.87  & 0.71  & 0.38  & 0.82  & 0.74  & 0.70  & 0.70  & 0.66  & 0.81  \\
    0.96  & 0.94  & 0.96  & 0.88  & 0.95  & 0.93  & 0.96  & 0.94  & 0.55  & 1.00  & 0.84  & 0.97  & 0.82  & 0.84  & 0.95  & 0.88  & 0.95  & 0.97  & 0.97  & 0.96  & 0.94  \\
    0.84  & 0.94  & 0.89  & 0.99  & 0.77  & 0.92  & 0.80  & 0.94  & 0.81  & 0.84  & 1.00  & 0.93  & 0.99  & 0.99  & 0.69  & 0.99  & 0.94  & 0.92  & 0.92  & 0.93  & 0.91  \\
    0.98  & 0.99  & 0.97  & 0.96  & 0.90  & 0.94  & 0.92  & 0.99  & 0.66  & 0.97  & 0.93  & 1.00  & 0.91  & 0.94  & 0.87  & 0.96  & 0.98  & 0.96  & 0.96  & 0.99  & 0.96  \\
    0.80  & 0.91  & 0.89  & 0.98  & 0.74  & 0.89  & 0.77  & 0.91  & 0.87  & 0.82  & 0.99  & 0.91  & 1.00  & 0.96  & 0.68  & 0.98  & 0.94  & 0.91  & 0.91  & 0.91  & 0.92  \\
    0.88  & 0.96  & 0.88  & 0.98  & 0.78  & 0.93  & 0.81  & 0.96  & 0.71  & 0.84  & 0.99  & 0.94  & 0.96  & 1.00  & 0.69  & 0.98  & 0.93  & 0.90  & 0.90  & 0.93  & 0.88  \\
    0.85  & 0.84  & 0.87  & 0.73  & 0.97  & 0.82  & 0.95  & 0.84  & 0.38  & 0.95  & 0.69  & 0.87  & 0.68  & 0.69  & 1.00  & 0.73  & 0.87  & 0.93  & 0.93  & 0.87  & 0.85  \\
    0.89  & 0.95  & 0.93  & 1.00  & 0.77  & 0.96  & 0.80  & 0.95  & 0.82  & 0.88  & 0.99  & 0.96  & 0.98  & 0.98  & 0.73  & 1.00  & 0.96  & 0.93  & 0.93  & 0.95  & 0.95  \\
    0.93  & 0.97  & 0.99  & 0.96  & 0.89  & 0.90  & 0.91  & 0.97  & 0.74  & 0.95  & 0.94  & 0.98  & 0.94  & 0.93  & 0.87  & 0.96  & 1.00  & 0.96  & 0.96  & 0.99  & 0.97  \\
    0.90  & 0.94  & 0.93  & 0.93  & 0.93  & 0.95  & 0.94  & 0.94  & 0.70  & 0.97  & 0.92  & 0.96  & 0.91  & 0.90  & 0.93  & 0.93  & 0.96  & 1.00  & 1.00  & 0.95  & 0.97  \\
    0.90  & 0.94  & 0.93  & 0.93  & 0.93  & 0.95  & 0.94  & 0.94  & 0.70  & 0.97  & 0.92  & 0.96  & 0.91  & 0.90  & 0.93  & 0.93  & 0.96  & 1.00  & 1.00  & 0.95  & 0.97  \\
    0.96  & 0.99  & 0.98  & 0.95  & 0.90  & 0.89  & 0.93  & 0.99  & 0.66  & 0.96  & 0.93  & 0.99  & 0.91  & 0.93  & 0.87  & 0.95  & 0.99  & 0.95  & 0.95  & 1.00  & 0.94  \\
    0.89  & 0.91  & 0.96  & 0.95  & 0.84  & 0.94  & 0.86  & 0.91  & 0.81  & 0.94  & 0.91  & 0.96  & 0.92  & 0.88  & 0.85  & 0.95  & 0.97  & 0.97  & 0.97  & 0.94  & 1.00  \\
\end{array}
\right]
}
\end{equation}


%

\begin{figure}
    \centering
    \subfigure[The similar nodes of node 1]{
    \label{similar_1} 
    \centering
    \includegraphics[scale=0.3]{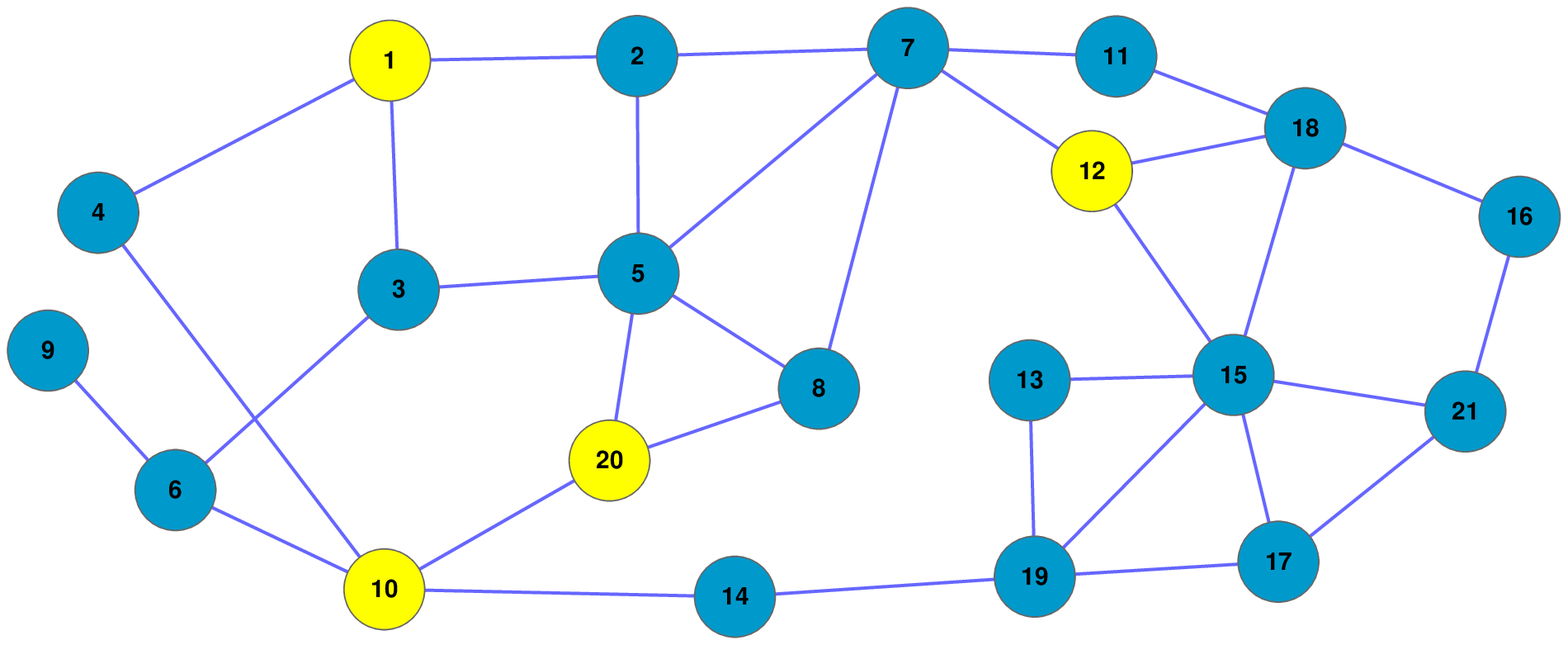}}
    \hspace{0.5cm}
    \subfigure[The similar nodes of node 2]{
    \label{similar_2} 
    \centering
    \includegraphics[scale=0.3]{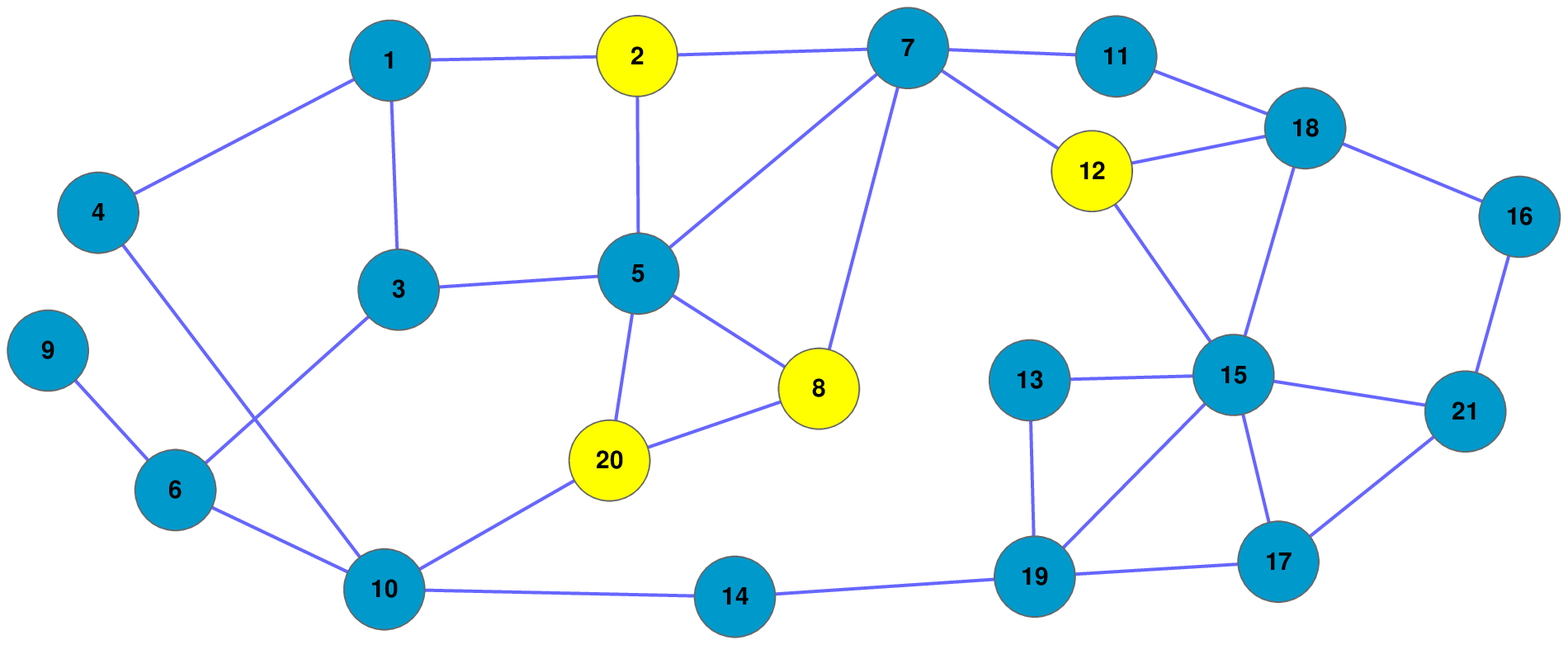}}
    \hspace{0.5cm}
    \subfigure[The similar nodes of node 3]{
    \label{similar_3} 
    \centering
    \includegraphics[scale=0.3]{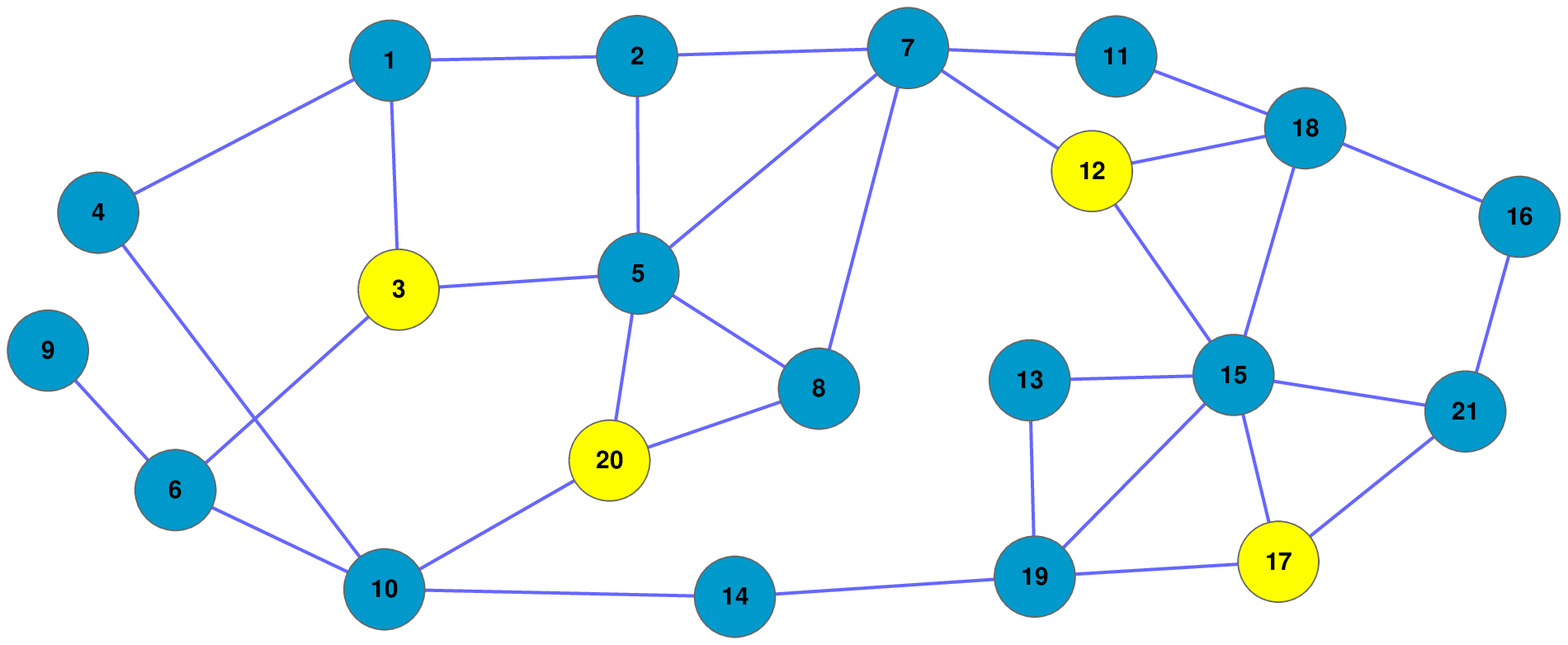}}
    \hspace{0.5cm}
    \subfigure[The similar nodes of node 4]{
    \label{similar_4} 
    \centering
    \includegraphics[scale=0.3]{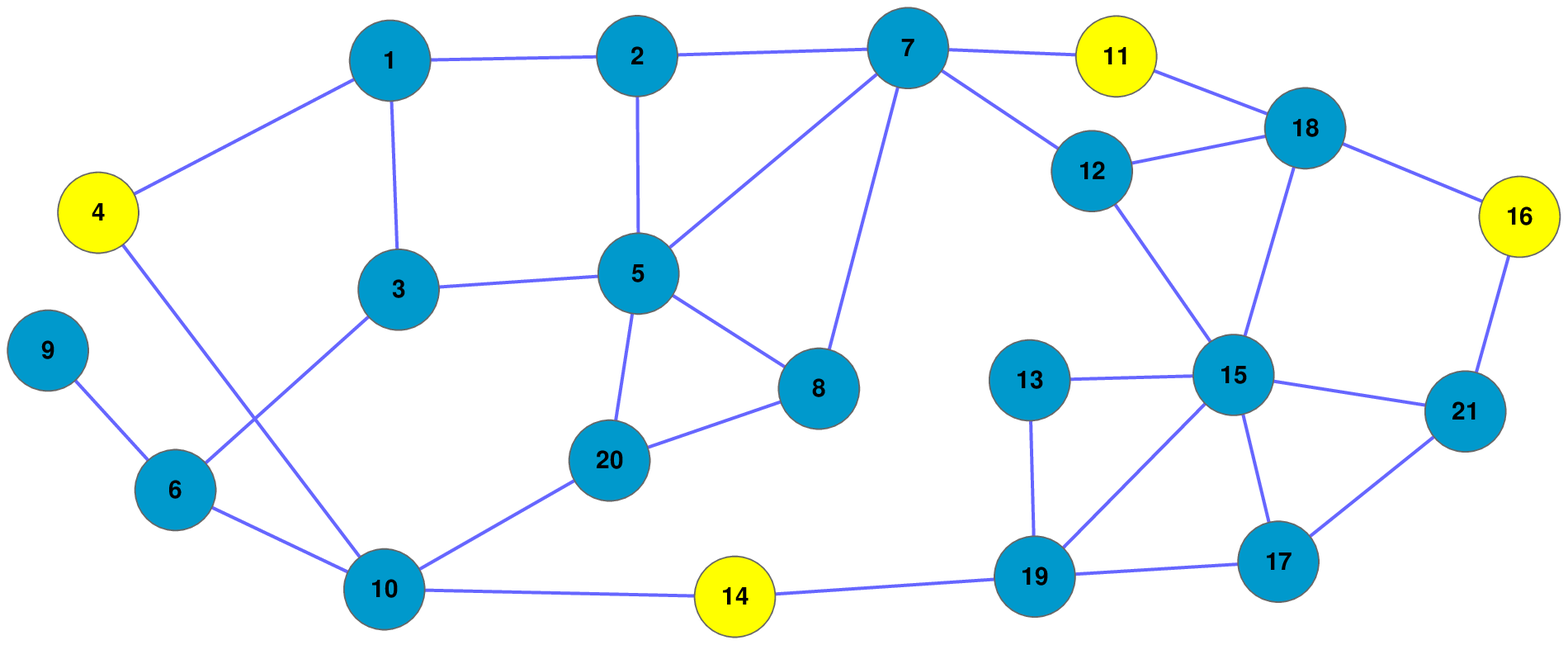}}
    \hspace{0.5cm}
    \subfigure[The similar nodes of node 5]{
    \label{similar_5} 
    \centering
    \includegraphics[scale=0.3]{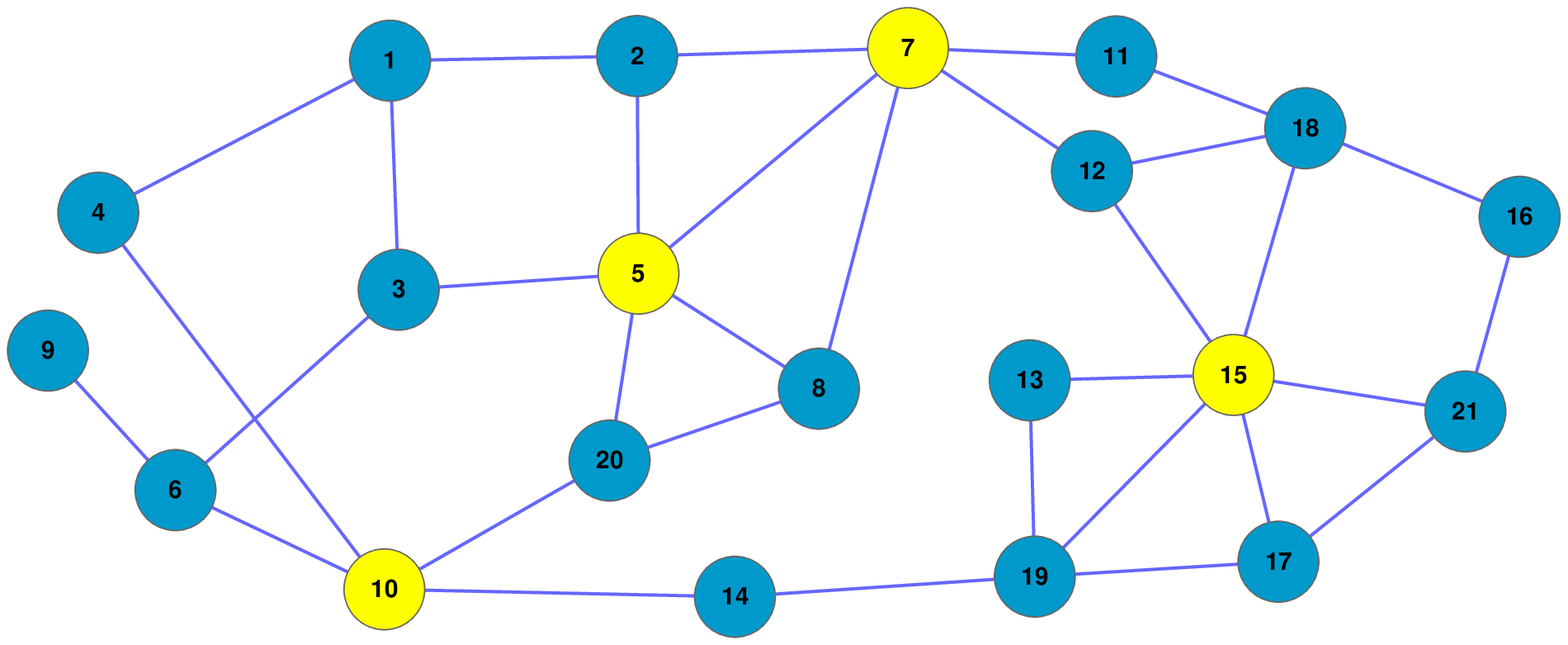}}
    \hspace{0.5cm}
    \subfigure[The similar nodes of node 6]{
    \label{similar_6} 
    \centering
    \includegraphics[scale=0.3]{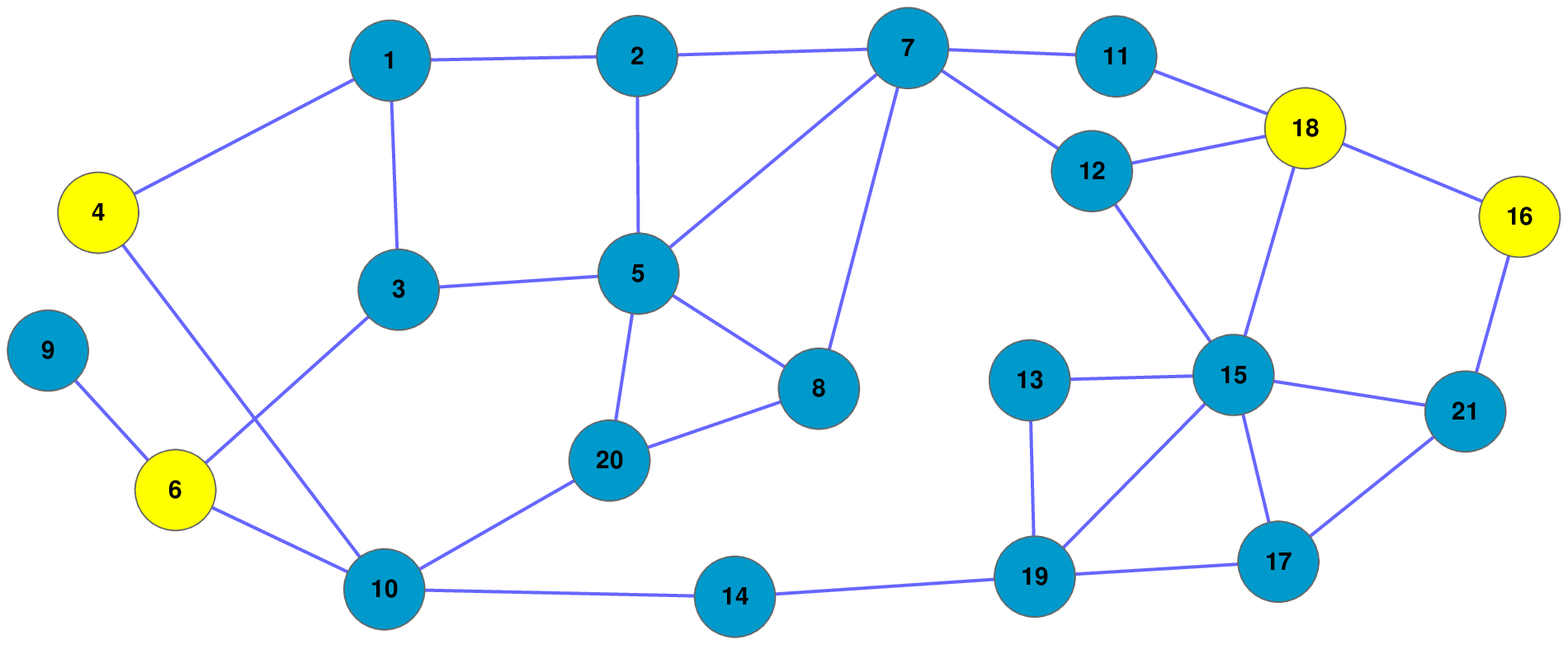}}
    \hspace{0.5cm}
    \subfigure[The similar nodes of node 7]{
    \label{similar_7} 
    \centering
    \includegraphics[scale=0.3]{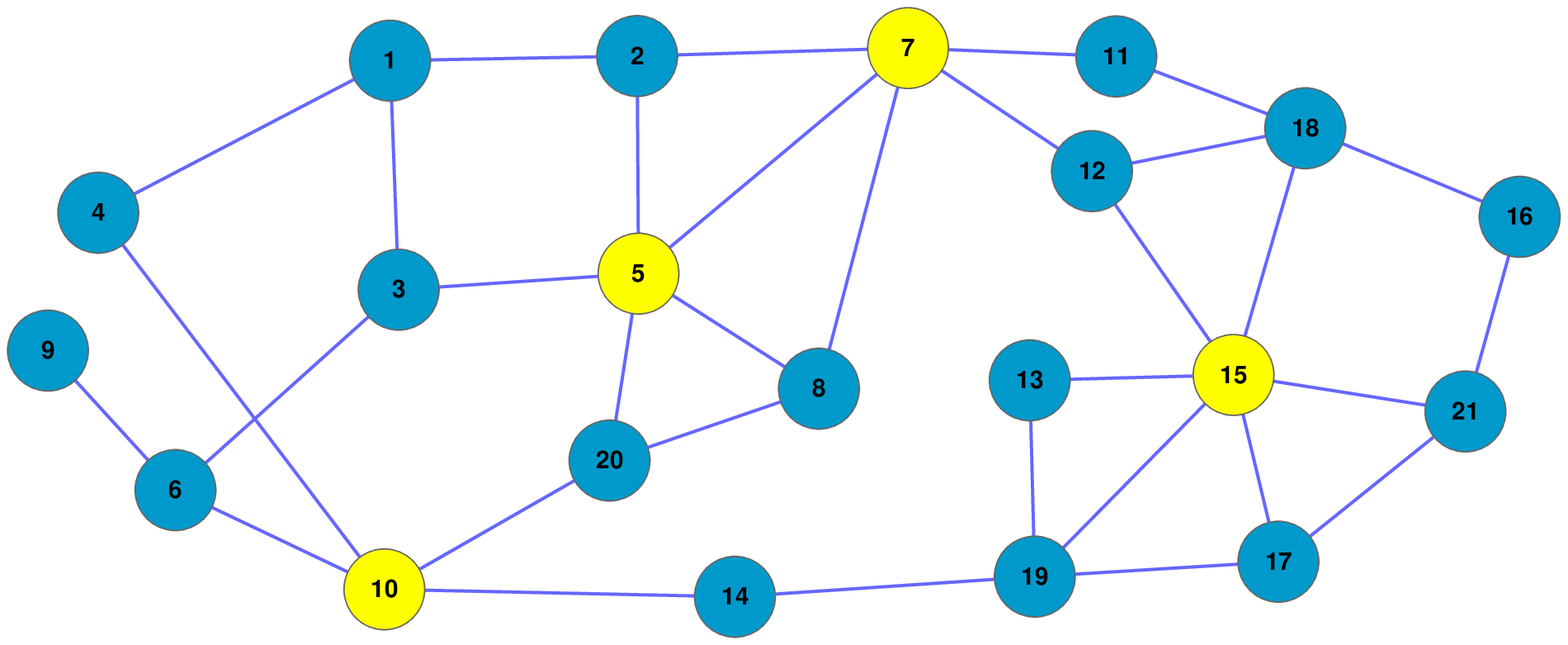}}
    \hspace{0.5cm}
    \subfigure[The similar nodes of node 8]{
    \label{similar_8} 
    \centering
    \includegraphics[scale=0.3]{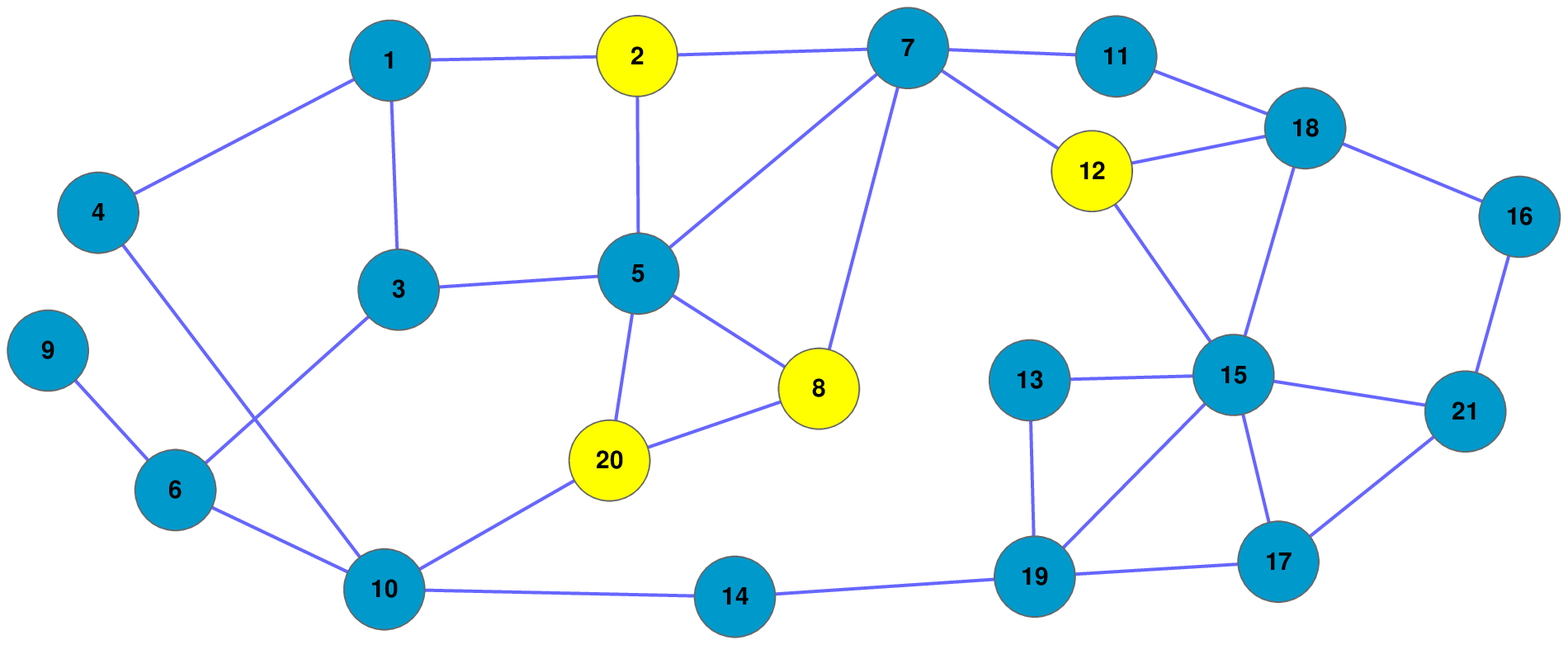}}
    \hspace{0.5cm}
    \subfigure[The similar nodes of node 9]{
    \label{similar_9} 
    \centering
    \includegraphics[scale=0.3]{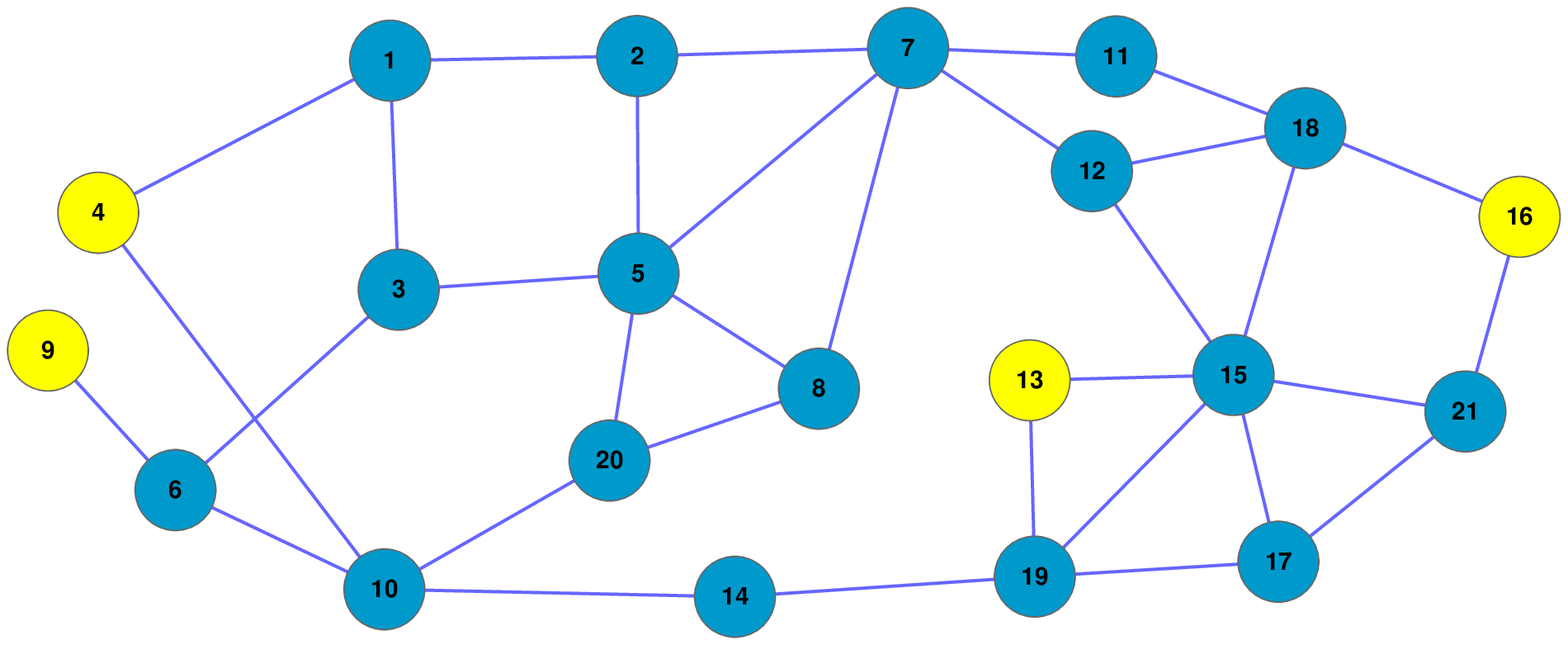}}
    \hspace{0.5cm}
    \subfigure[The similar nodes of node 10]{
    \label{similar_10} 
    \centering
    \includegraphics[scale=0.3]{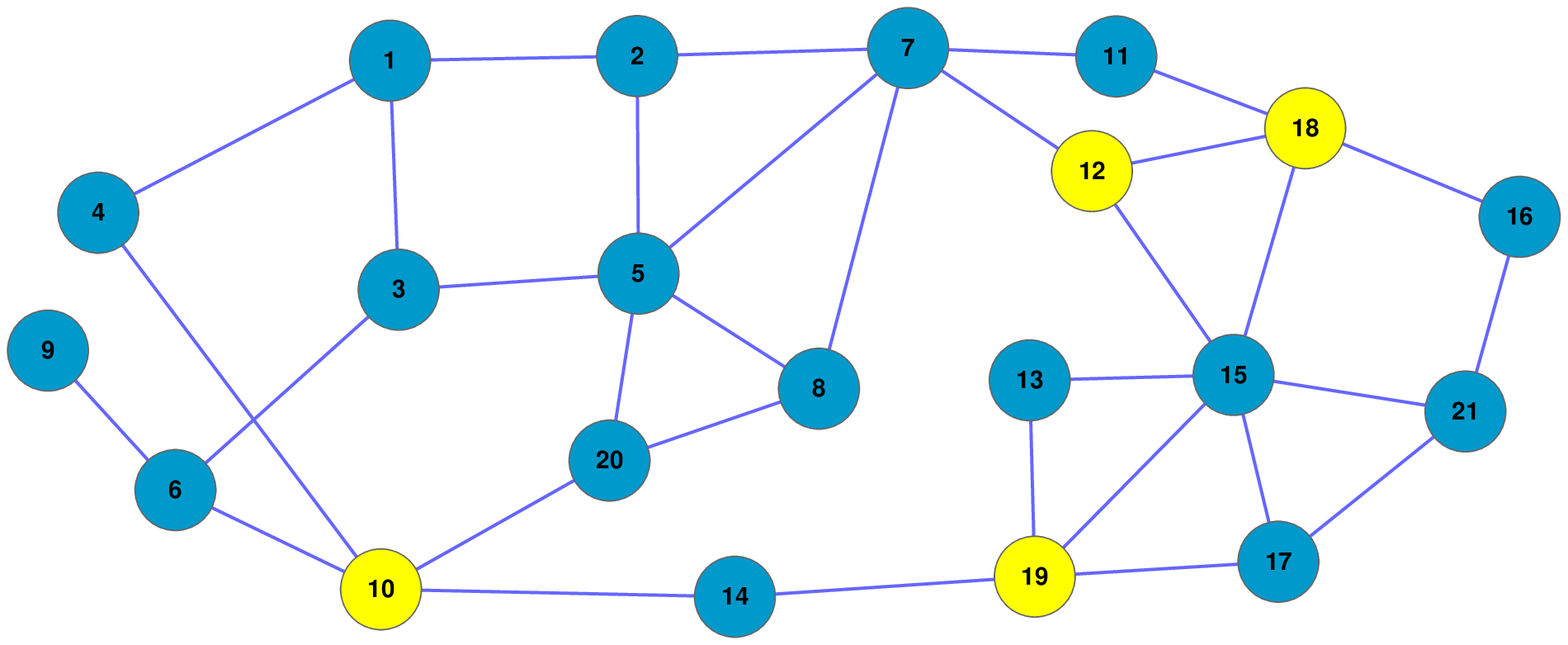}}
  \caption{The similar nodes of each node (From node 1 to node 10) }
  \label{Similar node 1_10}
\end{figure}

\begin{figure}
    \centering
    \subfigure[The similar nodes of node 11]{
    \label{similar_11} 
    \centering
    \includegraphics[scale=0.3]{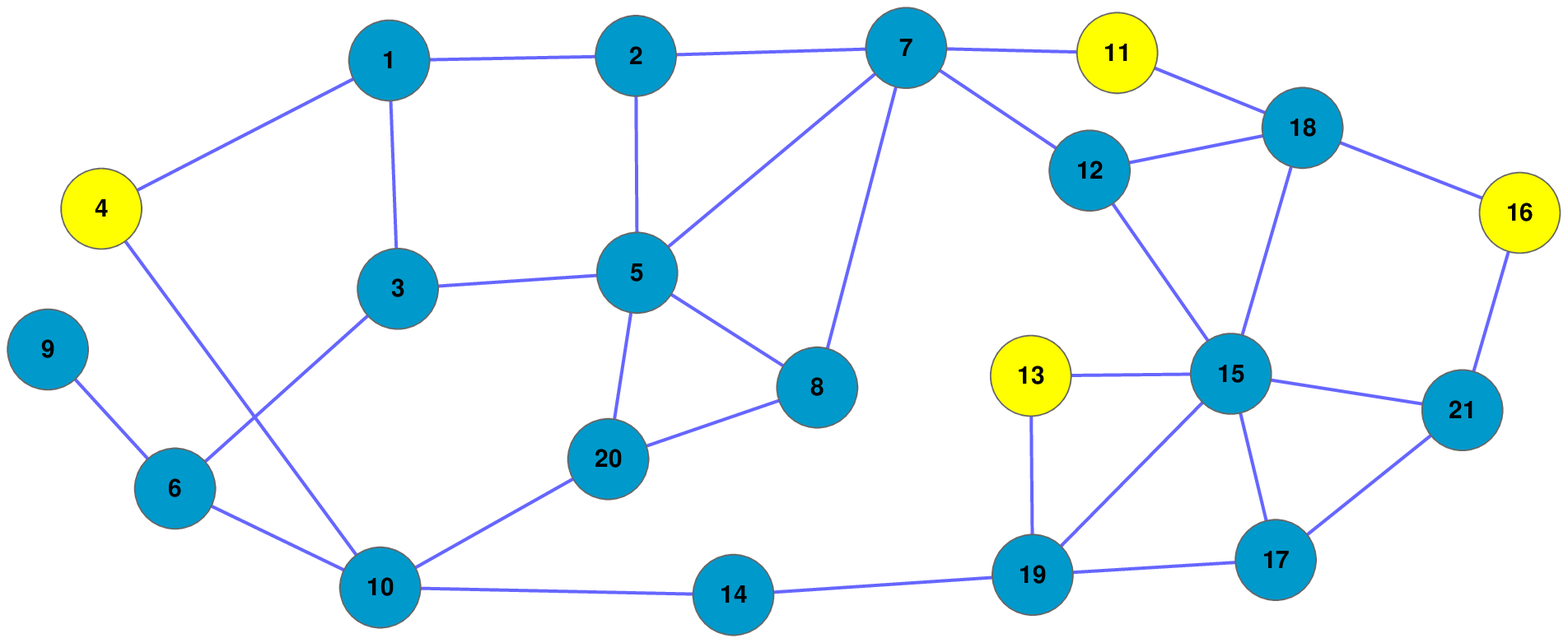}}
    \hspace{0.5cm}
    \subfigure[The similar nodes of node 12]{
    \label{similar_12} 
    \centering
    \includegraphics[scale=0.3]{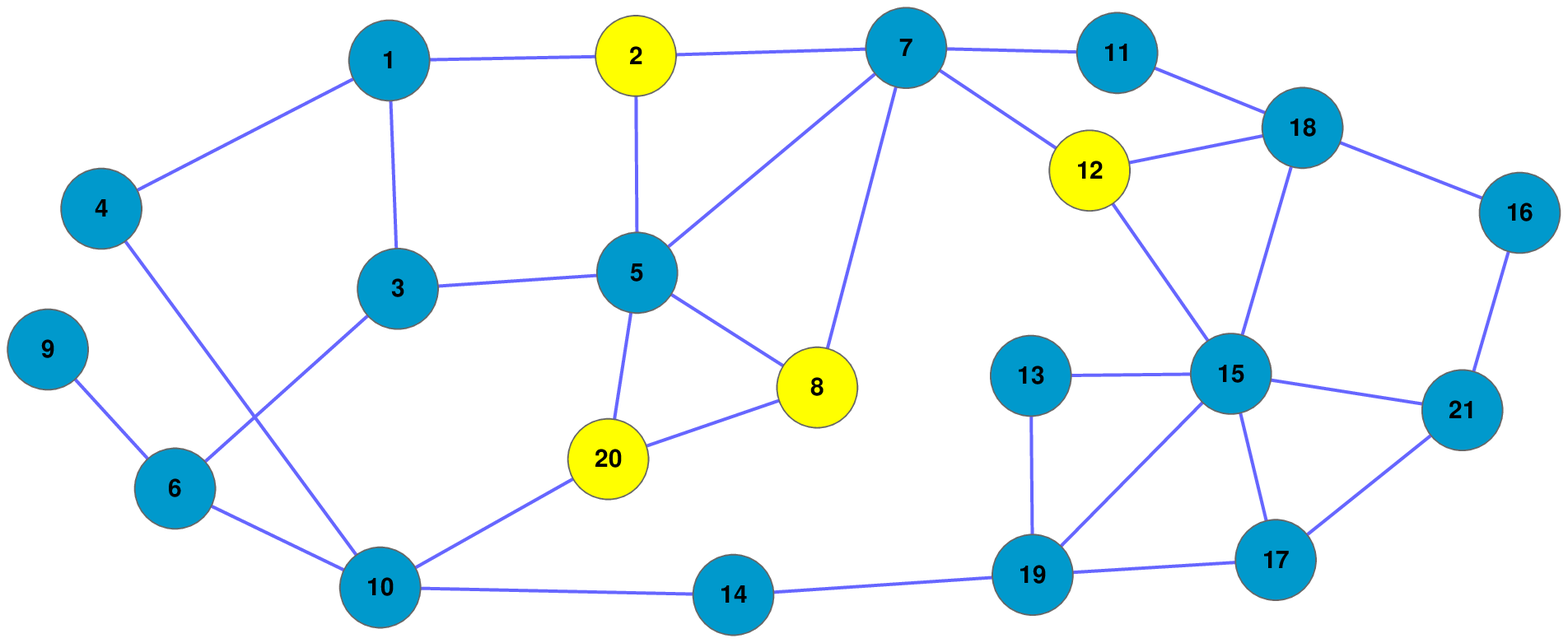}}
    \hspace{0.5cm}
    \subfigure[The similar nodes of node 13]{
    \label{similar_13} 
    \centering
    \includegraphics[scale=0.3]{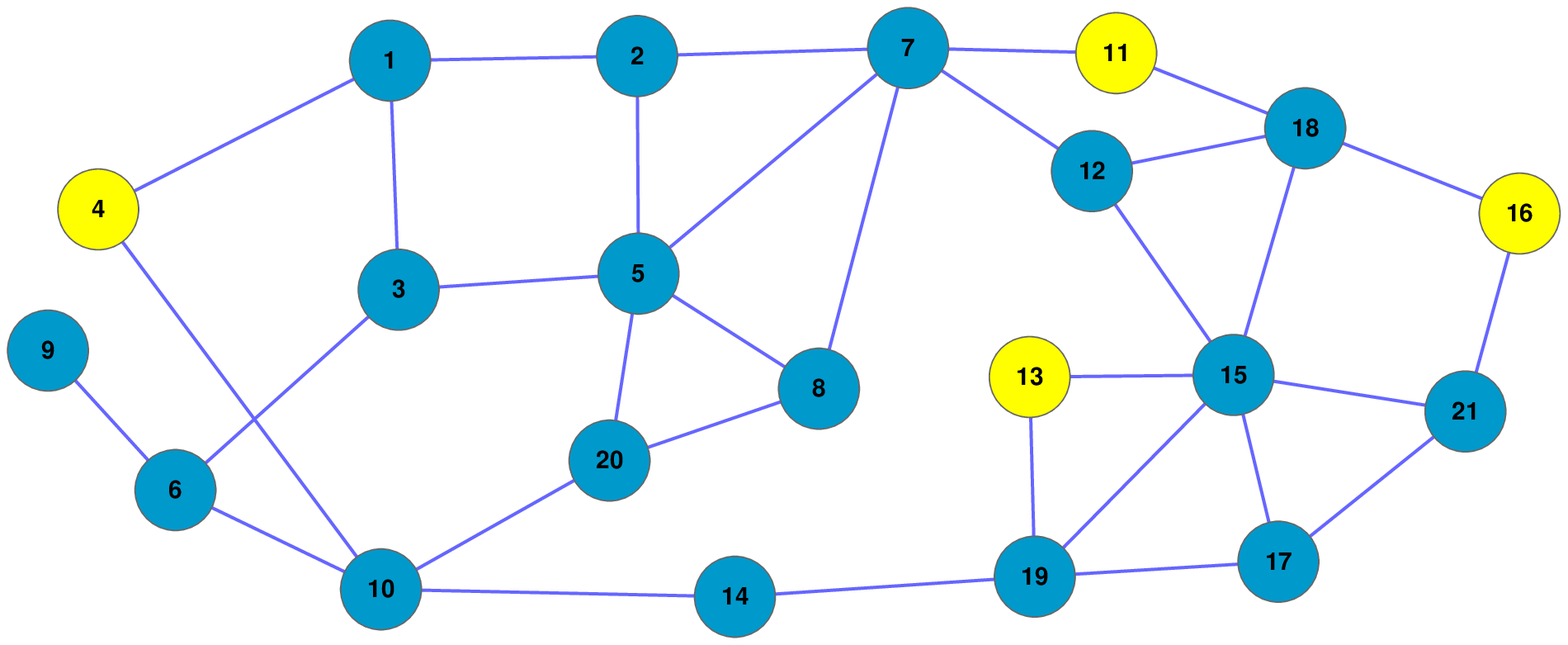}}
    \hspace{0.5cm}
    \subfigure[The similar nodes of node 14]{
    \label{similar_14} 
    \centering
    \includegraphics[scale=0.3]{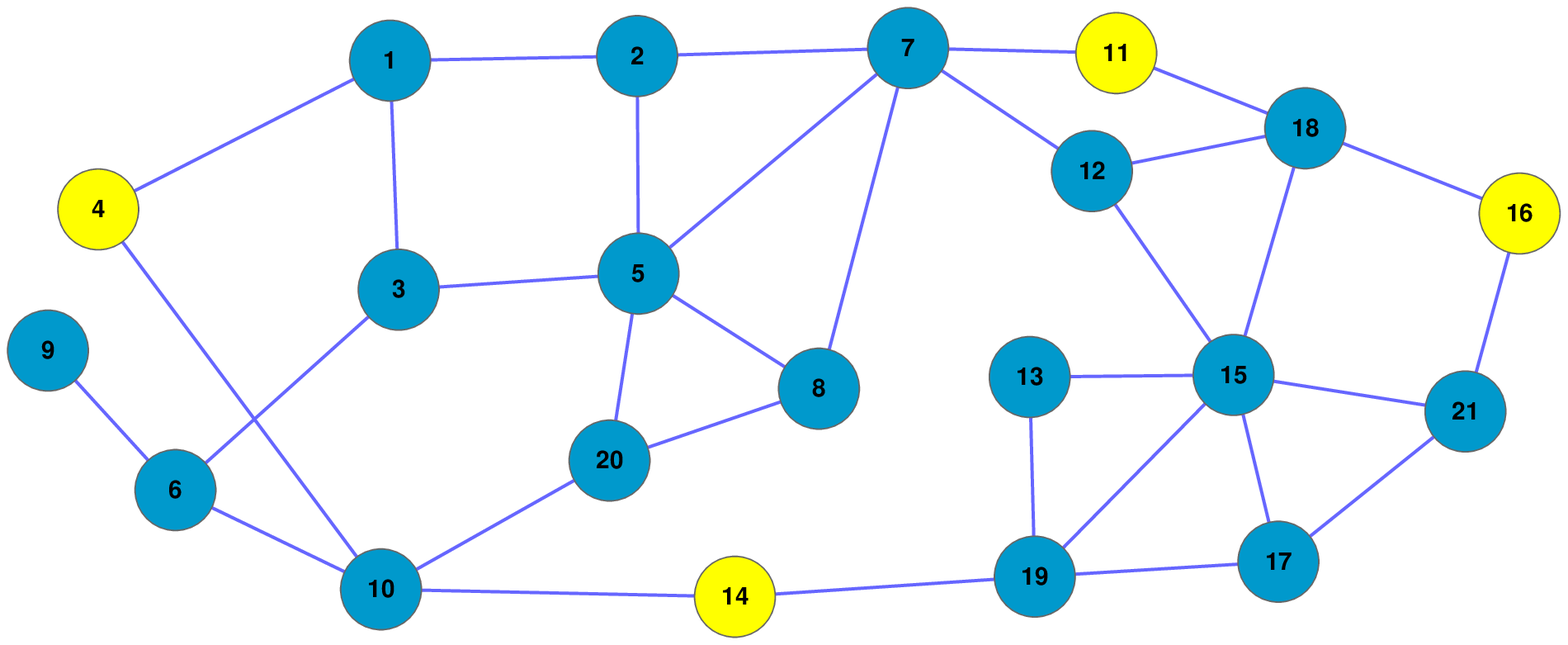}}
    \hspace{0.5cm}
    \subfigure[The similar nodes of node 15]{
    \label{similar_15} 
    \centering
    \includegraphics[scale=0.3]{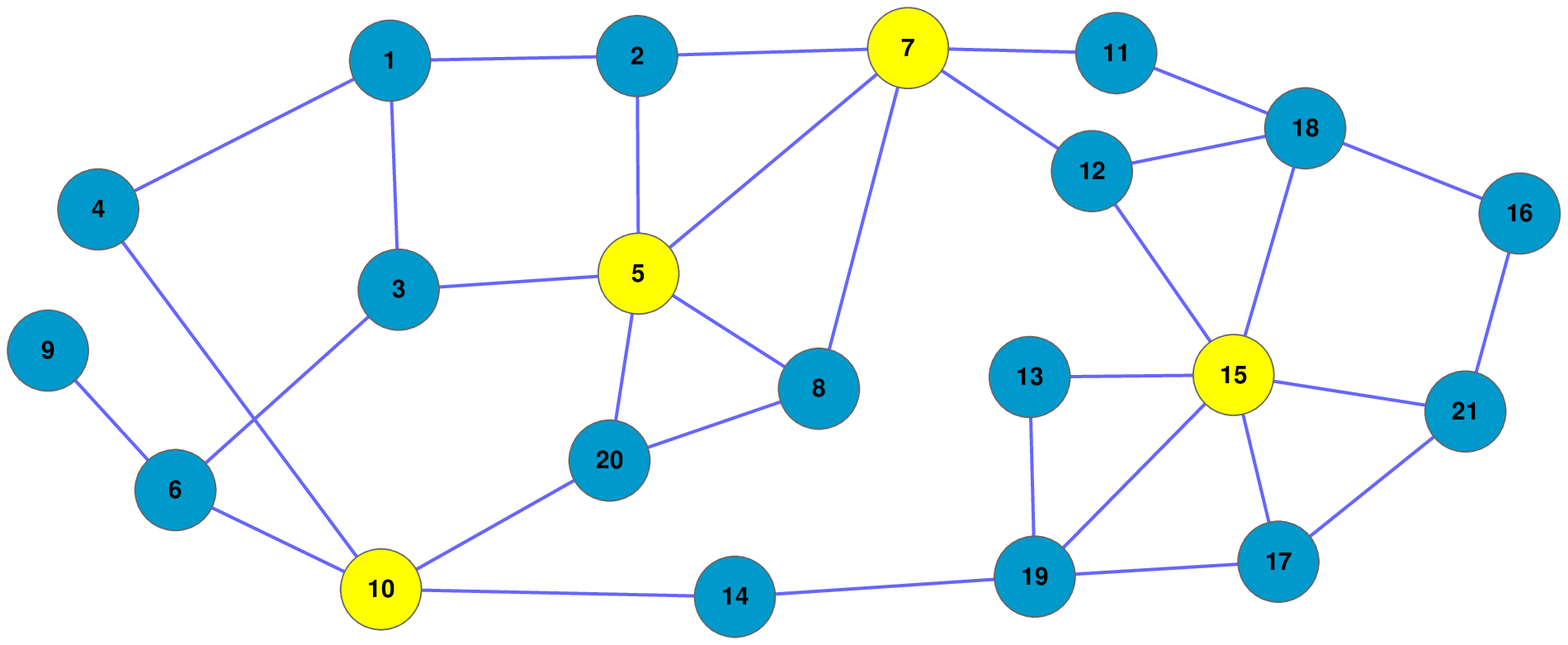}}
    \hspace{0.5cm}
    \subfigure[The similar nodes of node 16]{
    \label{similar_16} 
    \centering
    \includegraphics[scale=0.3]{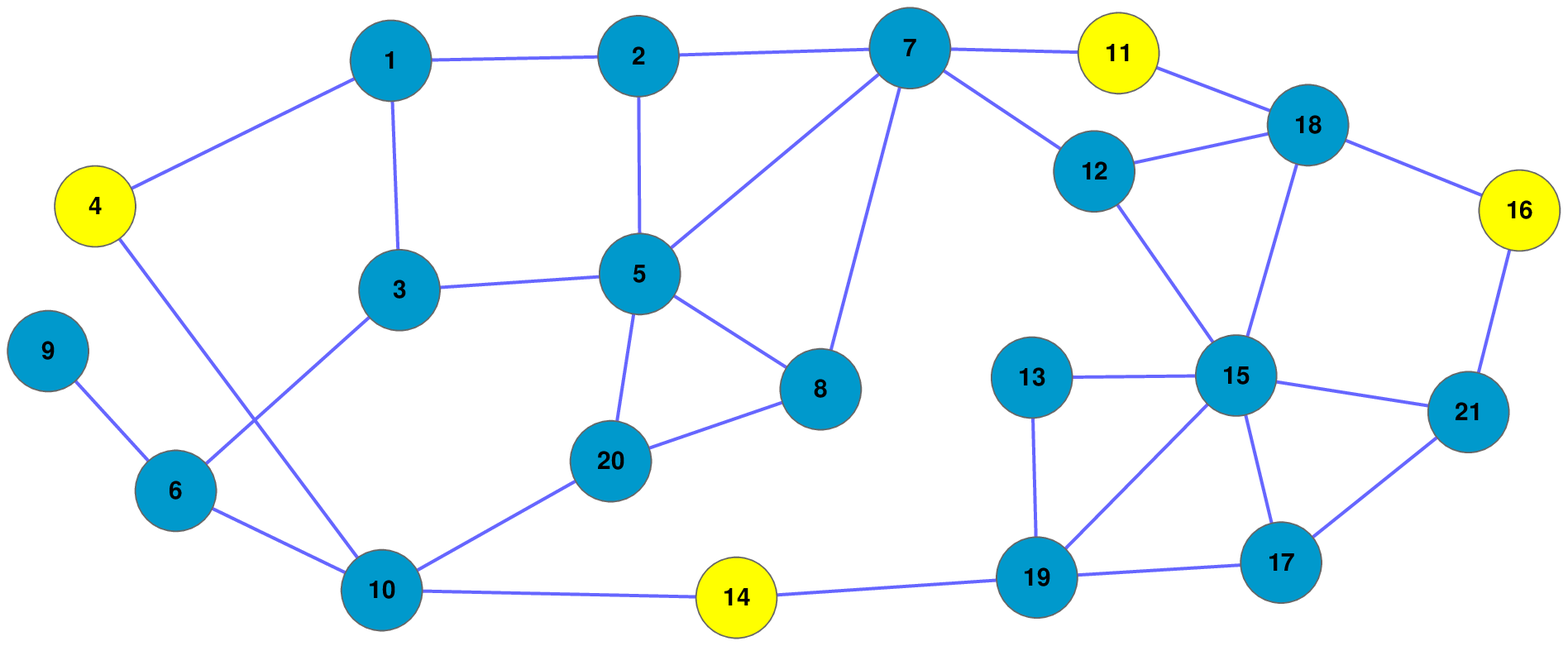}}
    \hspace{0.5cm}
    \subfigure[The similar nodes of node 17]{
    \label{similar_17} 
    \centering
    \includegraphics[scale=0.3]{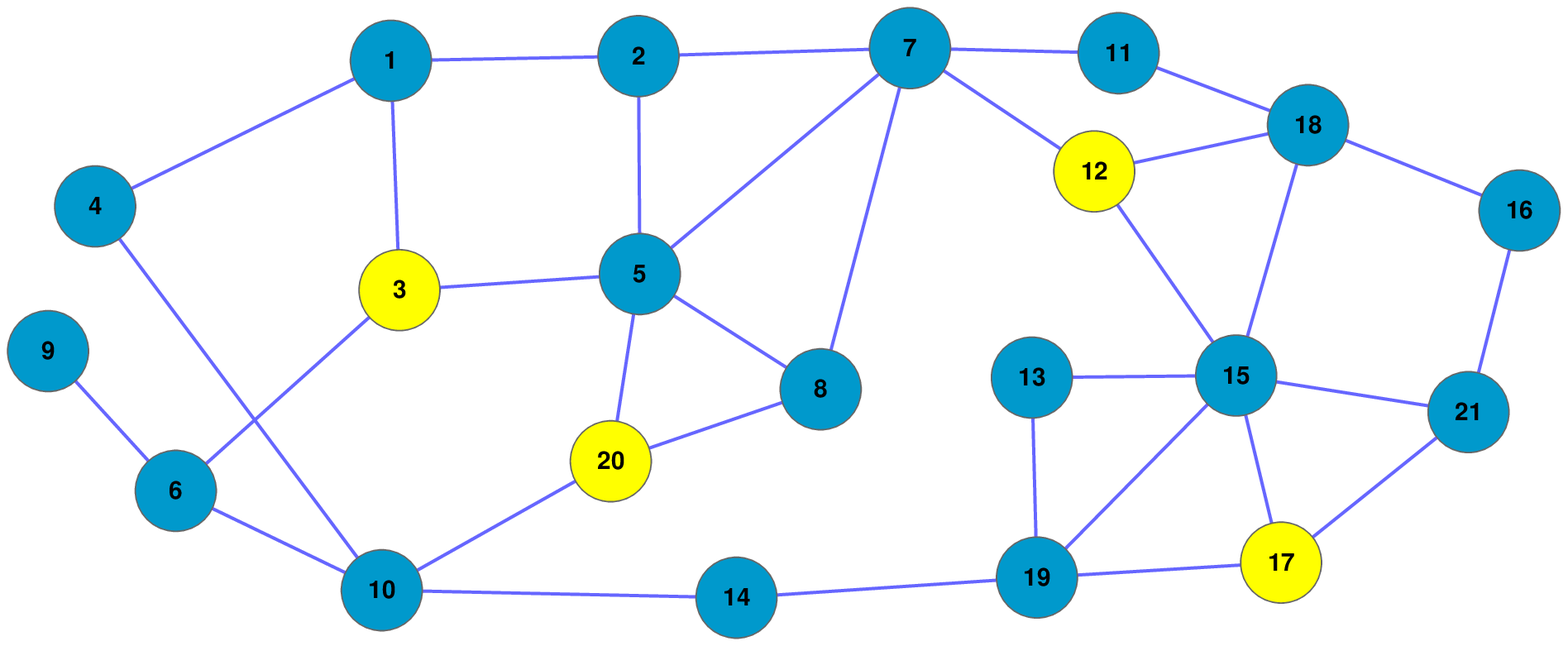}}
    \hspace{0.5cm}
    \subfigure[The similar nodes of node 18]{
    \label{similar_18} 
    \centering
    \includegraphics[scale=0.3]{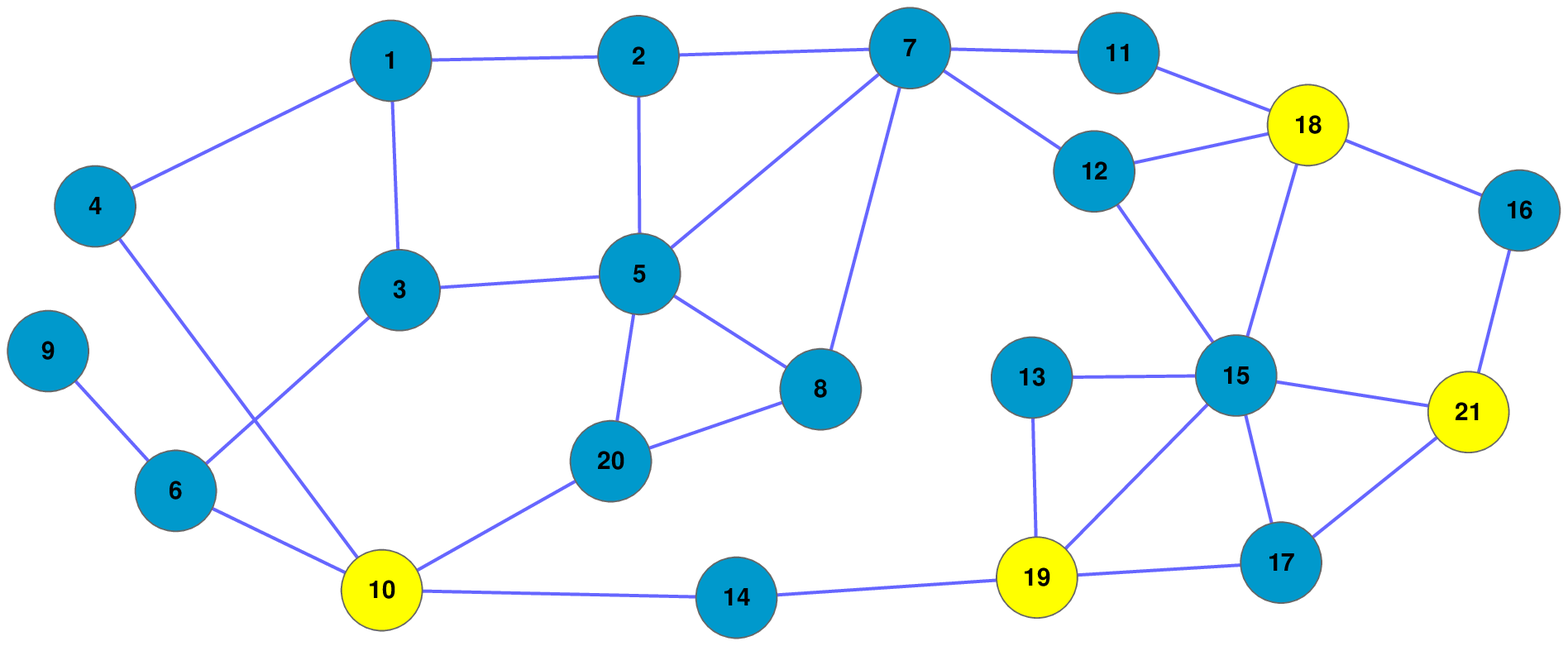}}
    \hspace{0.5cm}
    \subfigure[The similar nodes of node 19]{
    \label{similar_19} 
    \centering
    \includegraphics[scale=0.3]{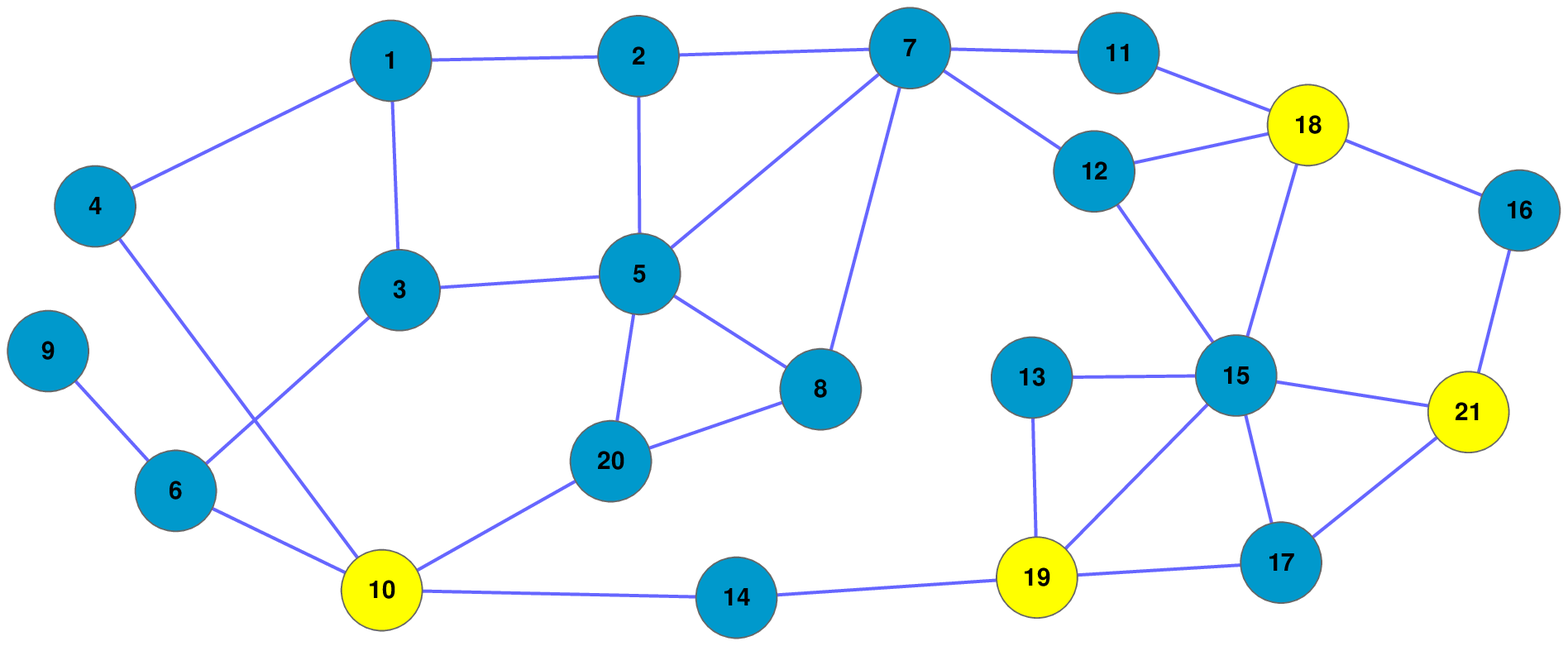}}
    \hspace{0.5cm}
    \subfigure[The similar nodes of node 20]{
    \label{similar_20} 
    \centering
    \includegraphics[scale=0.3]{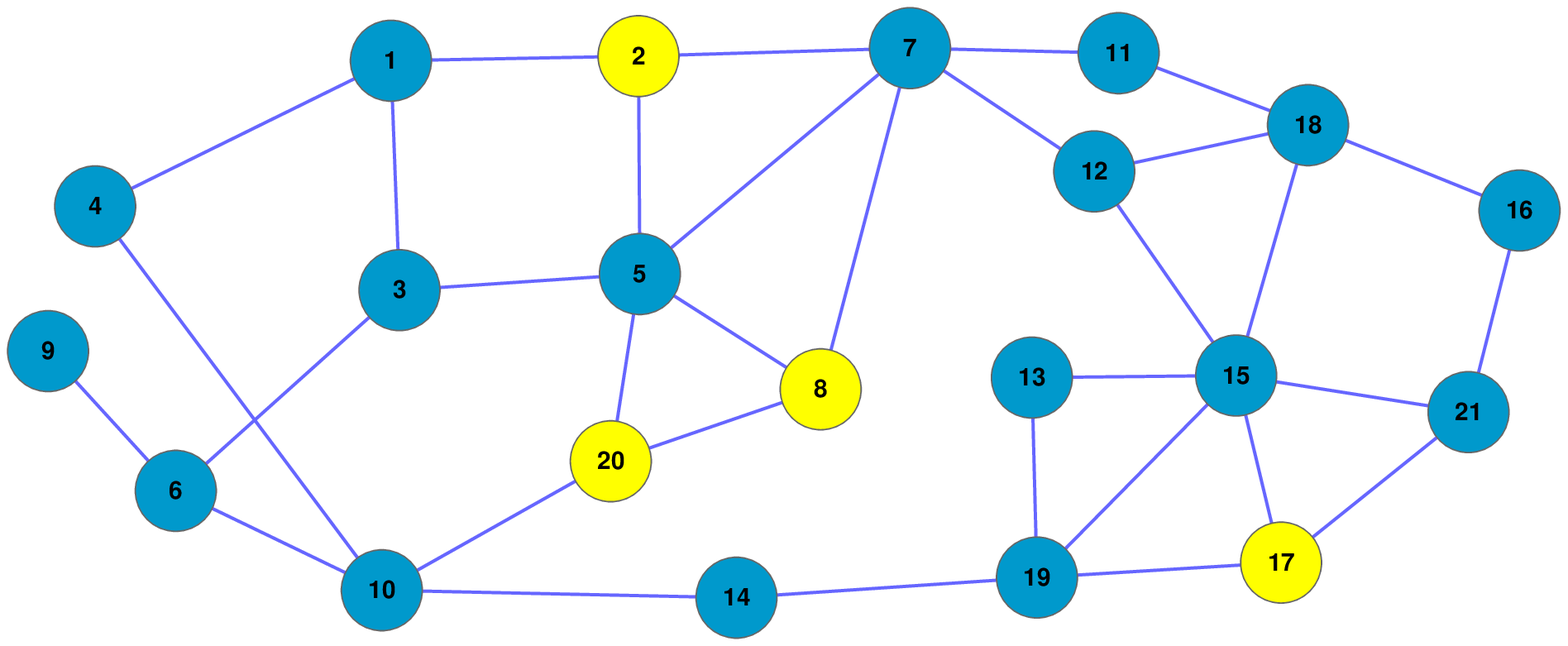}}
  \caption{The similar nodes of each node (From node 11 to node 20) }
  \label{Similar node 11_20}
\end{figure}

From the similarity matrix, we can find that the node 2 and node 8, the node 4 and node 16, the node 18 and node 19 have the same structure in the example network (Network A-21). The details is shown in the Fig. \ref{SimilarHHHnode}:

\begin{figure}
    \centering
    \subfigure[The details of node 2 and node 8]{
    \label{A21_2_8} 
    \centering
    \includegraphics[scale=0.5]{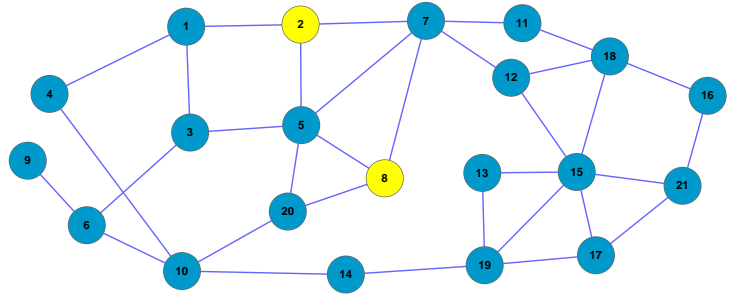}}
    \hspace{0.5cm}
    \subfigure[The details of node 4 and node 16]{
    \label{A21_4_16} 
    \centering
    \includegraphics[scale=0.5]{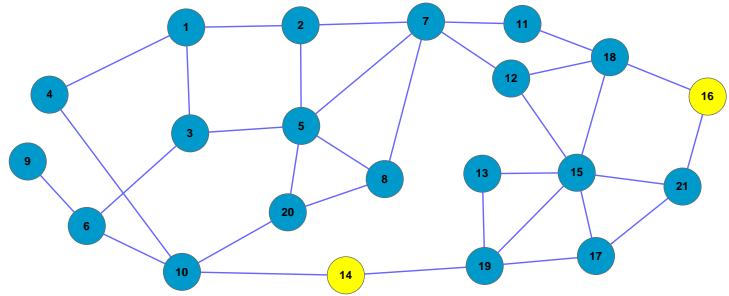}}
    \hspace{0.5cm}
    \subfigure[The details of node 18 and node 19]{
    \label{A21_18_19} 
    \centering
    \includegraphics[scale=0.5]{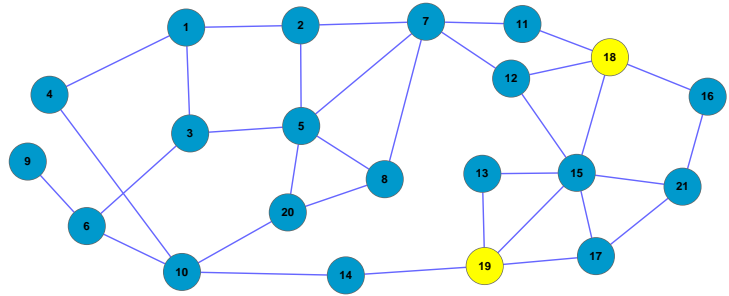}}
  \caption{The details of those nodes which have same structure. }
  \label{SimilarHHHnode}
\end{figure}

From the similarity matrix, we also have find that the node 9 has the lowest similarity to others nodes and the node 12 have the highest similarity to others nodes.

From the results of our test on the example network (Network A-21), the measurement of the similarity of the nodes based on the Relative-entropy is an reasonable and useful method. The method also can be use to node classify in the complex networks. The node 12 have the highest similarity to others nodes. The degree of node 12 is equal to 3. In the example network (Network A-21) most node's degree is equal to 3. It shows from the other hands that the degree is very important to describe the structure property of the complex networks. The node 9 is a marginal node, because this is no node has a high similarity to it.
\section{Application}
\label{application}
In the section, the new method is used to find the most similar node in four real networks. The four networks are the Zachary's Karate Club network (Karate) \cite{uci}, the US-airport network (Us-airport) \cite{networkdata}, Email networks (Email) \cite{networkdata}and the Germany highway networks (Highway) \cite{nettt}. The results are shown as follows:

\begin{table}[htbp]
  \centering
  \caption{The most similar node and the most marginal node in the four real networks  }
    \begin{tabular}{ccccc}
    \toprule
     Network     & Nodes & Edages & High similarity node & Low similarity node \\
    \midrule
    Karate & 34    & 78    & 28    & 12 \\
    Us-airport & 332   & 2126  & 55    & 118 \\
    Email & 1133  & 10902 & 855   & 644 \\
    Highway & 1168  & 2481  & 31    & 798 \\
    \bottomrule
    \end{tabular}%
  \label{tab:addlabel}%
\end{table}%

\section{Conclusion}
\label{conclusion}
Measure the similarity of the node in the complex networks is an interesting topic. In this paper, a new method which is based on the Relative-entropy is proposed the measure the similarity of the nodes in the complex networks. The nodes with common structure have a high similarity to others. When the similarity between those nodes is equal to 1, it means that those two nodes have same structure property in the complex networks. The nodes which have influential to other or the nodes which are marginal nodes in the complex networks have a low similarity to others. The results in this paper show that, the proposed methods is useful and reasonable to measure the similarity of the node in the complex networks.
\section*{Acknowledgments}
The work is partially supported by National Natural Science Foundation of China (Grant No. 61174022), Specialized Research Fund for the Doctoral Program of Higher Education (Grant No. 20131102130002), R$\&$D Program of China (2012BAH07B01), National High Technology Research and Development Program of China (863 Program) (Grant No. 2013AA013801), the open funding project of State Key Laboratory of Virtual Reality Technology and Systems, Beihang University (Grant No.BUAA-VR-14KF-02). Fundamental Research Funds for the Central Universities No. XDJK2015D009. Chongqing Graduate Student Research Innovation Project (Grant No. CYS14062)

%



\bibliographystyle{elsarticle-num}
\bibliography{zqreference}






\end{document}